\documentclass[10pt,prl,aps,twocolumn,showpacs]{revtex4}
\usepackage{graphicx}

\usepackage{amssymb}

\begin{document}

\title{Z$_{\rm 4}$--U(1) crossover of the order parameter symmetry \\ 
in a two-dimensional valence-bond-solid}

\author{Jie Lou}
\affiliation{Department of Physics, Boston University, 
590 Commonwealth Avenue, Boston, Massachusetts 02215}

\author{Anders W. Sandvik}
\affiliation{Department of Physics, Boston University, 
590 Commonwealth Avenue, Boston, Massachusetts 02215}

\begin{abstract}
We discuss ground-state projector simulations of a modified two-dimensional $S=1/2$ Heisenberg model in the valence bonds basis. 
Tuning matrix elements corresponding to the diagonal and off-diagonal terms in the quantum dimer model, we show that there is a quantum 
phase transition from the antiferromagnet into a columnar valence-bond-solid (VBS). There are no signs of discontinuities, suggesting 
a continuous or very weakly first-order transition. The Z$_4$-symmetric VBS order parameter exhibits an emergent U$(1)$ symmetry as 
the phase transition is approached. We extract the associated length-scale governing the U$(1)$--Z$_4$ cross-over 
inside the VBS phase.
\end{abstract}

\date{\today}


\maketitle

A valence-bond-solid (VBS) is a magnetically disordered state of a quantum spin system in which translational symmetry is spontaneously 
broken due to the formation of a pattern of strong and weak bond correlations $\langle {\bf S}_i \cdot {\bf S}_j\rangle$ (where $i,j$ are
nearest-neighbor sites). Using an SU$(N)$ generalization of the Heisenberg model, Read and Sachdev showed that a four-fold degenerate columnar VBS 
ground state can be expected on the square lattice \cite{read}. Numerical studies have found evidence for such VBS states in frustrated SU$(2)$
symmetric systems \cite{j1j2}, but because of technical limitations, in particular the sign problem in quantum Monte Carlo (QMC) 
simulations \cite{henelius}, the nature of the strongly-frustrated ground state remains controversial \cite{isaev}. Another challenging issue 
is how the ground state evolves from an antiferromagnet (AF) into a VBS. According to the ``Landau rules'', one would expect a direct 
transition  between these states to be first-order \cite{sirker}, because unrelated symmetries are broken. There could also be an intervening disordered 
(spin liquid) phase \cite{anderson} or a coexistence region. Senthil {\it et al.}~recently suggested an alternative scenario for a
generic continuous transition based on a ``deconfined'' quantum critical point (DQCP) associated with spinon deconfinement \cite{senthil,nogueira}. 
This proposal has generated significant interest, as well as controversy. An extended ``J-Q'' Heisenberg model has been introduced \cite{sandvik1} 
which is not frustrated, in the standard sense, but includes a four-spin interaction which destroys the AF order and leads to a VBS ground state. 
This model is amenable to large scale QMC studies, which show scaling behavior consistent with a DQCP \cite{sandvik1,melko}. Other studies 
dispute these findings, however \cite{jiang}. Numerical studies of the proposed field theory describing the deconfined quantum critical point 
are also subject to conflicting interpretations \cite{motrunich,kuklov}. Further studies of AF--VBS transitions is thus called for.

In this Letter we address an important aspect of the VBS state and the AF--VBS transition, namely, the nature of the quantum fluctuations of the VBS 
order parameter. In the DQCP theory, the Z$_4$ symmetric lattice-imposed structure of the VBS is a dangerously irrelevant, and, as a consequence, U$(1)$ 
symmetry emerges close to the DQCP \cite{senthil}. An U$(1)$ symmetric VBS order parameter was indeed confirmed in the studies of the J-Q model 
\cite{sandvik1,melko,jiang}, and also in simulations of the SU$(N)$ Heisenberg model with $N > 4$ \cite{kawashima}. However, the expected cross-over into 
a Z$_4$ symmetric distribution inside the VBS phase was not observed. This can be interpreted as the lattice sizes studied so far being smaller than 
the spinon confinement length-scale $\Lambda$, which governs the U$(1)$--Z$_4$ cross-over \cite{lou1}. $\Lambda$ should diverge as $\xi_d^a$, where $\xi_d$ 
is the dimer (VBS) correlation length and $a>1$ \cite{senthil}, and for a finite lattice with $L \ll \Lambda$ the distribution should be U$(1)$ symmetric. 
The models studied so far have a rather weak VBS order, and hence $\xi_d$ is large, which likely makes it difficult to satisfy $L \gg \xi_d^a$. The exponent 
$a$ is not known. 

Here we introduce a way to generate much more robust VBS states, with which we can study the U$(1)$--Z$_4$ cross-over already on small lattices.  Our 
approach is based on a ground-state projector QMC method operating in the valence bond (VB) basis \cite{sandvik2,sandvik3,liang2}. Starting from some trial 
state $|\Psi\rangle$, the ground state of a hamiltonian $H$ can be obtained by applying a high power of $H$; $|\Psi_0\rangle \sim H^m|\Psi\rangle$. Consider 
the $S=\frac{1}{2}$ Heisenberg model written as a sum of singlet projection operators $H_{ij}$,
\begin{equation}
H = -J\sum_{\langle i,j\rangle} H_{ij},~~~~H_{ij}=\hbox{$\frac{1}{4}$}-{\bf S}_i \cdot {\bf S}_j,
\end{equation}
where $\langle i,j\rangle$ denotes nearest neighbors on a square lattice of $N=L^2$ sites. In the VB basis the trial state $|\Psi\rangle$ is a superposition of 
singlet products $|(a_1,b_1)\cdots(a_{N/2},b_{N/2})\rangle$, where $(a,b)=(\uparrow_{a}\downarrow_{b}-\downarrow_{a}\uparrow_{b})/\sqrt{2}$ with $a$ and $b$ sites 
on different sublattices. We here use the amplitude-product state of Liang {\it et al.} \cite{liang1,lou2}. A singlet projector can have two different effects upon 
acting on a VB state;
\begin{eqnarray}
&& H_{ab}|\cdots (a,b)(c,d)\cdots \rangle = 1|\cdots (a,b)(c,d)\cdots \rangle \label{vbdia} \\
&& H_{ad}|\cdots (a,b)(c,d)\cdots \rangle = \hbox{$\frac{1}{2}$}|\cdots (a,d)(c,b)\cdots \rangle.  \label{vboff}~~ ~~~~~
\end{eqnarray}
These rules form the basis of the VB projector method \cite{sandvik3,liang2}, where $H^m$ is expanded in its strings 
of $m$ singlet projectors. Expectation values of operators $O$ are obtained by importance-sampling the VBs and operator strings produced when
expanding 
\begin{equation}
\langle O \rangle = \frac{\langle \Psi|H^m OH^m|\Psi\rangle}{\langle \Psi|H^mH^m|\Psi\rangle}.
\end{equation}
For details of these procedures we refer to Refs.~\cite{sandvik2,sandvik3}. 

\begin{figure}
\includegraphics[width=4.5cm,clip]{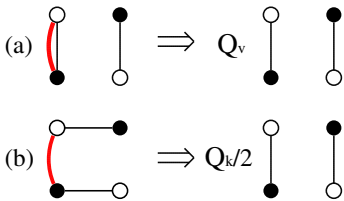}
\caption{(Color on-line) Diagonal (a) and off-diagonal (b) singlet projection operations (indicated by the arches) on VB pairs on a plaquette. The matrix elements 
(\ref{vbdia}) and (\ref{vboff}) corresponding to these operations are multiplied by $Q_v$ and $Q_k$, respectively. In (a), the factor is $2Q_v$ if there is a VB 
also on to the left side of the operator. For all other bond configurations the matrix elements remain those in (\ref{vbdia}) and (\ref{vboff}).}
\label{fig1}
\vskip-3mm
\end{figure}

In Ref.~\cite{sandvik1}, the J-Q model, which includes a four-spin coupling consisting of terms $-QH_{ij}H_{kl}$, with $ij$ and $kl$ site pairs on two opposite edges 
of a plauqtte, was studied using the VB projector method. The $Q$ term naturally favors singlet formation on plaquettes, and this was shown to lead to a VBS state 
when $J/Q<(J/Q)_c$, $(J/Q)_c \approx  0.04$. Here we introduce another mechanism leading to VBS formation. We define an effective hamiltonian based on the Heisenberg 
model in the VB basis, by changing the diagonal matrix element $1$ in Eq.~(\ref{vbdia}) and the off-diagonal matrix element $\frac{1}{2}$ in Eq.~(\ref{vboff}) 
to $Q_v$ and $\frac{1}{2}Q_k$, respectively, for $H_{ab}$ acting on VBs on opposite edges of the same plaquette. 
These operations, illustrated in Fig.~\ref{fig1}, correspond to the kinetic- and potential-energy terms of the quantum 
dimer model \cite{rokhsar}. There, however, the Hilbert space consists of only dimers connecting nearest-neighbor sites, whereas we here keep 
the full space of VBs connecting any pair of sites on different sublattices. In the quantum dimer model, the dimer configurations are also 
considered as orthogonal states, whereas we here keep the singlet nature of the VBs, whence the states are non-orthogonal. The non-orthogonality 
may at first sight seem problematic, because when $Q_v,Q_k \not =1$ the hamiltonian is non-hermitean. We therefore refer to it as an {\it pseudo hamiltonian} 
in the VB basis. However, in spite of this, the states generated by the projection procedure (with the sampling weights modified by the presence of the factors 
$Q_v$ and $Q_k$, and taking the power $m$ large enough for convergence to the $m=\infty$ limit) are completely well-defined SU$(2)$ invariant quantum states. 
We can thus think of the modified projection technique as a means of generating a family of states parametrized by $Q_v$ and $Q_k$. Moreover, 
there must be some corresponding hamiltonians, defined in terms of the standard spin operators ${\bf S}_i$, which have these states as their ground states. 
Although we are not able to write down these hamiltonians (which likely contain multi-spin interactions, possibly long-ranged), it is still useful to study 
the evolution of the states as a function of $Q_v$ and $Q_k$. Here we will consider two cases; $Q_v \ge 1, Q_k=1$ and $Q_k \ge 1, Q_v=1$, 
which we refer to as the $Q_v$ and $Q_k$ models, respectively. Both these models indeed undergo AF--VBS transitions.

\begin{figure}
\centerline{\includegraphics[width=7.25cm,clip]{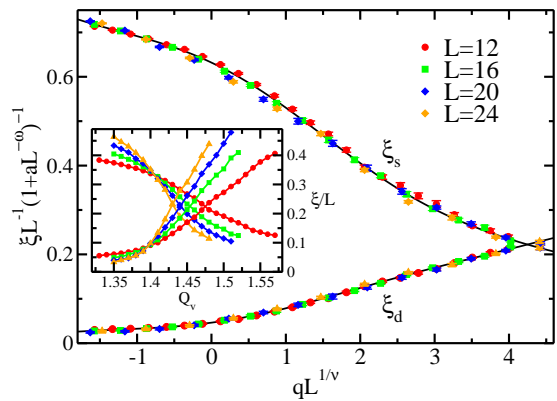}}
\caption{(Color on-line) Finite-size scaling of the spin and dimer correlation lengths. The inset shows $\xi/L$ versus 
the coupling, with crossing points tending toward $Q_v^c \approx 1.40$.}
\label{xi}
\vskip-3mm
\end{figure}

We calculate the square of the staggered magnetization, $M^2=\langle {\bf M} \cdot {\bf M}\rangle$, where 
\begin{equation}
{\bf M}=\frac{1}{N}\sum_{x,y}(-1)^{x+y} {\bf S}_{x,y}
\end{equation}
is the operator for the AF order parameter. The columnar VBS operator for $x$-oriented bonds is
\begin{eqnarray}
D_x=\frac{1}{N}\sum_{x,y}(-1)^{x} {\bf S}_{x,y} \cdot {\bf S}_{x+1,y},
\end{eqnarray}
and $D_y$ is defined analogously. We calculate the squared order parameter, $D^2=\langle D_x^2+D_y^2\rangle$, and the distribution $P(D_x,D_y)$ as in \cite{sandvik1}. 
Results for these quantities and the corresponding spin and dimer correlation lengths $\xi_s$ (spin) and $\xi_d$ (defined through the momentum-space 
second moments of the spin and dimer correlation functions) indicate coinciding critical points for the AF and VBS order parameters. In the following we first discuss the finite 
size scaling behavior of the $Q_v$ model.

\begin{figure}
\centerline{\includegraphics[width=6.5cm,clip]{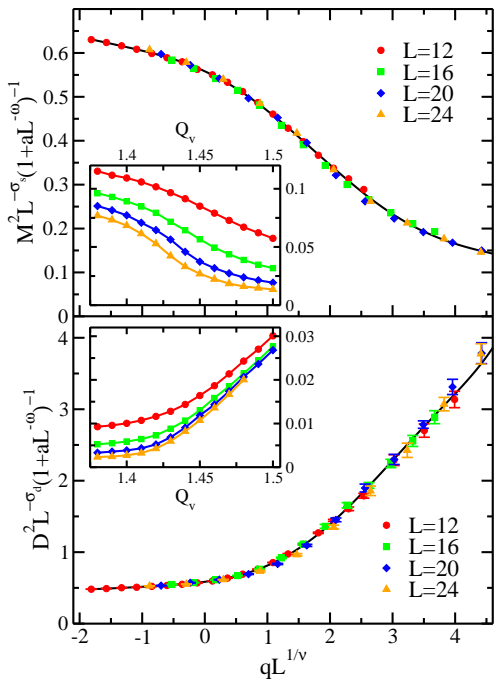}}
\caption{(Color on-line) Finite size scaling of the squared AF (top panel) and VBS (bottom panel) order parameters of the $Q_v$ model. 
The insets show the unscaled data.}
\label{MD}
\vskip-3mm
\end{figure}

We define a reduced coupling $q=Q_v-Q^c_v$. Then, if there indeed is a single critical point, there is AF order for $q<0$ and VBS order for $q>0$, and in the 
thermodynamic limit the squared spin and dimer order parameters should scale as $M^2 \sim (-q)^{2\beta_s}$ and $D^2 \sim q^{2\beta_d}$ inside the respective 
phases. To extract $Q_v^c$ and the exponents, we use standard finite-size scaling forms,
\begin{eqnarray}
M^2              &=&  L^{\sigma_s}(1+aL^{- \omega})F_s(qL^{1/\nu}),  \\
D^2              &=&  L^{\sigma_d}(1+aL^{- \omega})F_d(qL^{1/\nu}),   \\
\xi_{s,d}        &=&  L(1+aL^{- \omega}) G_{s,d}(qL^{1/\nu}).
\end{eqnarray}
where $\sigma_s=2\beta_s/\nu$, $\sigma_d=2\beta_d/\nu$ and the correlation length exponent $\nu$ is the same for all the quantities (as required in the
DQCP theory). The scaling functions $F_{s,d}$ and $G_{s,d}$ are extracted in the standard way by adjusting the critical point and exponents to collapse 
finite-size data onto common curves. Since our lattices are not very large, $L \le 24$, a subleading correction helps significantly to scale the data. 
In all cases we find that $\omega=1$ works well (the prefactor $a$ is quantity-dependent, however).

Results are shown in Fig.~\ref{xi} and \ref{MD}. All the data can be scaled with $Q^c_v=1.400(5)$, $\nu=0.78(3)$, $\beta_s=0.27(2)$ and 
$\beta_d=0.68(3)$. Here $\nu$ and $\beta_d$ are approximately the same, within error bars, as in the J-Q model \cite{sandvik1}, while $\beta_s$ is very 
different---for the J-Q model $\beta_s \approx \beta_d=0.63(2)$ was found. The range of system sizes is quite small and we cannot, of course, exclude drifts in the exponents for larger lattices, nor a very weakly first-order transition. 

Turning to the $Q_k$ model, it is more demanding computationally, because the critical point is rather large, $Q_k= 2.5(1)$, leading to a lower acceptance rate in simulations 
close to the transition than for the $Q_v$ model. We can therefore not reach the same level of precision for the exponents. The results are nevertheless
consistent with a continuous transition and exponents similar to those of the $Q_v$ model.

\begin{figure}
\centerline{\includegraphics[width=5cm,clip]{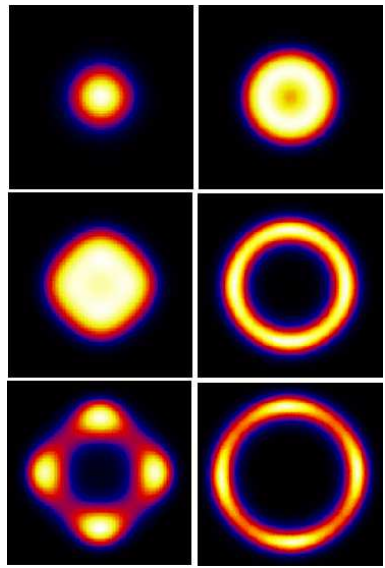}}
\caption{(Color on-line) VBS order=parameter distributions. The left column is for $Q_k=1,Q_v=1.44,1.48,1.54$ (from top) on $12 \times 12$ lattices.
The right column is for $Q_v=1,Q_k=2.5,3.5,5.0$ (from top) on $16\times 16$ lattices.}
\label{histo}
\vskip-3mm
\end{figure}

Unfortunately, we cannot easily calculate the dynamic exponent $z$ with the present approach, because it requires access to the triplet sector, e.g., to extract the 
spin gap $\Delta \sim L^{-z}$. Our model is explicitly defined only in the singlet sector. While one can extend the VB basis and the projection scheme to triplets 
\cite{sandvik2,sandvik3}, the extension of the $Q_v$ and $Q_k$ models to this sector is not unique, and $z$ may depend on how that is accomplished. We could in principle 
calculate gaps in the singlet sector, but this is much more complicated.

Our main interest in studying the $Q_v$ and $Q_k$ models is in the distribution $P(D_x,D_y)$ of the columnar dimer order parameter. While this is a basis
dependent quantity, it still provides direct information on the order parameter symmetry. In a columnar symmetry-broken VBS state, we expect a distribution with 
a single peak located on the $x$ or $y$ axis, while in a plaquette state the peak should be on one of the $45^\circ$ rotated axes. In simulations that do not break 
the symmetry, we expect four-fold symmetric distributions, with peak locations corresponding to the type of VBS as above. In previous studies of VBS states, only ring-shaped 
distributions were observed \cite{sandvik1,melko,jiang,kawashima}, however, which can be taken as a confirmation of the predicted \cite{senthil} emergent 
$U(1)$ symmetry close to a DQCP. One would then expect the four-fold symmetry to appear for large systems, $L \gg \Lambda$, inside the VBS phase, as has been 
observed explicitly in a classical XY model including dangerously irrelevant $Z_q$ ($q\ge 4$) perturbations \cite{lou1}. With the $Q_v$ and $Q_k$ models, 
we can reach further inside the VBS phases than in the previously studied quantum spin systems, and, as seen in Fig.~\ref{histo}, we can indeed follow the evolution from 
$U(1)$ to $Z_4$ symmetric distributions as a function of the coupling constants  even for modest system sizes. The peak locations correspond to columnar VBS states for 
both models, although the shapes of the distributions are qualitative different in other respects. The previously observed ring-shaped distributions \cite{sandvik1,kawashima}
are more reminicant of those for the $Q_k$ model.

\begin{figure}
\centerline{\includegraphics[width=7.25cm,clip]{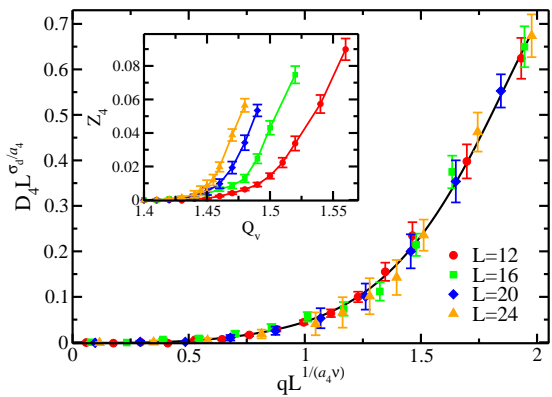}}
\caption{(Color on-line) Finite-size scaling of anisotropy order parameter. The inset shows the unscaled data.}
\label{z4}
\vskip-3mm
\end{figure}

To study the length scale $\Lambda$ which governs the $Z_4$--$U(1)$ crossover (and is related to the scaling dimension of a dangerously irrelevant field \cite{lou1}) 
we define an order parameter $D_4$ which is sensitive to the $Z_4$ anisotropy
\begin{eqnarray}
D_4&=&\int_{-1}^{1}dD_x\int_{-1}^{1}dD_yP(D_x,D_y) r_{xy}\cos(4\theta) \nonumber \\
   &=&\int_{0}^{1}dr\int_0^{2\pi}d\theta r^2P(r,\theta)\cos(4\theta),
\end{eqnarray}
where $r_{xy}=(D_x^2+D_y^2)^{1/2}$.
This order parameter should obey the finite-size scaling form  \cite{oshikawa,lou1};
\begin{equation}
D_4 = L^{\sigma_d/a_4}F_4(qL^{1/a_4\nu}),
\end{equation}
with $a_4>1$. The data can be scaled with $a=1.30(5)$, as shown in Fig.~\ref{z4} 
(where we use the same $Q^c_v,\sigma_d$, and $\nu$ as in Figs.~\ref{MD} and \ref{xi}). Here the error bars on the raw data, as seen in the inset, 
are much larger than for $D^2$, reflecting that slow angular fluctuations of the VBS order parameter in the simulations (which do not affect the rotationally-invariant $D^2$). 
In the classical XY model with $Z_4$ perturbation $a_4 \approx 1.1$ \cite{carmona,lou1}, and, thus, the $Q_v$ model has more prominent angular fluctuations. For the $Q_k$ model, we 
find an even larger $a_4 \approx 1.5$, although the error bars are rather large and we cannot say for sure that it is different from the $Q_v$ model.

To summarize, by tuning specific matrix elements in valence-bond QMC simulations, we are able to study a family of SU$(2)$ symmetric states undergoing AF--VBS phase transitions. 
Unlike previous studies of quantum spin models with VBS states, we are able to observe both the $Z_4$ symmetry of the order-parameter distribution (which arises from the nature of 
the VBS on the square lattice) deep inside the VBS phase and the cross-over into an emergent $U(1)$ symmetry upon approaching the transition point. We extracted the length scale 
$\Lambda$, which is associated with spinon deconfinement in the DQCP theory. While the correlation length exponent $\nu$ and the VBS order-parameter exponent $\beta_d$ are roughly consistent with 
those found previously for the J-Q model \cite{sandvik1,melko} (which is the best candidate so far for a DQCP), the AF exponent $\beta_s$ is significantly smaller. It thus appears 
that these transitions are in different universality classes. This implies that emergent U$(1)$ symmetry (which is associated with a dangerously irrelevant lattice-imposed potential) 
is a salient feature of VBS states that is more generic than the particular DQCP scenario by Senthil {\it al.}~\cite{senthil}. Our results then also point to a broader range of continuous 
(or, possibly, very weakly first-order) AF--VBS transitions.
 
We would like to thank Naoki Kawashima and Masaki Oshikawa for stimulating discussions. This work was supported by the NSF under grant No.~DMR-0803510. 

\null

\end{document}